# Analysis of azimuthal phase mask coronagraphs

François Hénault
Institut de Planétologie et d'Astrophysique de Grenoble
Université Grenoble-Alpes, Centre National de la Recherche Scientifique
B.P. 53, 38041 Grenoble – France

**ABSTRACT**

In this paper is presented an analytical study of the azimuthal phase-mask coronagraph currently envisioned for detecting and characterizing extra-solar planets. Special emphasis is put on the physical and geometrical interpretation of the mathematical development. Two necessary conditions are defined for achieving full extinction in the pupil plane of the coronagraph, stating that the complex amplitude generated by the phase mask should have zero average, on the one hand, and its Fourier coefficients should only be even, on the other hand. Examples of such phase functions are reviewed, including optical vortices, four-quadrant phase masks, and azimuthal cosine phase functions. Hints for building more sophisticated functions are also given. Finally, a simplified expression of light leaks due to mask imperfection is proposed.

## 1   INTRODUCTION

With the recent commissioning of ground instruments such as SPHERE [1] or GPI [2] and future space observatories like WFIRST-AFTA [3], coronagraphy should probably become an efficient tool for characterizing the atmospheres of extra-solar planets in the forthcoming years, especially when operated from space. Coronagraphic instruments are themselves divided into three or four families, one of which being Phase-mask coronagraphs (PMC) whose principle is depicted in Figure 1. It basically consists in setting transparent phase plates at the telescope focal plane or another optically conjugated plane. Such plates are spreading the optical beam onto a secondary pupil image, where an aperture stop (named Lyot stop) blocks the diffracted starlight. Very deep extinction ratios can then be achieved in the final image plane, revealing the presence of faint planets orbiting around the central star. Historically, the first phase function proposed by Roddier and Roddier was axis symmetric [4], but could not achieve perfect extinction over the full Lyot stop without apodizing the entrance pupil [5]. Nowadays the most popular phase masks are the Four-quadrant phase mask (FQPM) and vortex phase mask (VPM), whose phase functions only depend on an azimuthal coordinate $\theta$'. Extensive literature has already been published on their fundamental properties [6-9], manufacturing technologies [10-11], and experimental and observational achievements [12-13]. In particular, it has been demonstrated by using pure mathematical analysis [7-9, 14-15] that they produce full cancellation of the electric field over the whole surface of the Lyot stop, which is the basic requirement for this type of instruments. In spite of their rigor and clarity, these demonstrations essentially based on Hankel transforms may suffer from a lack of physical or geometrical interpretation. Moreover, they are essentially focused at the FQPM or VPM coronagraphs, though it can be conjectured that they are applicable to other types of azimuthal phase masks. The main goal of this paper is to extend the analysis to other phase masks and to provide engineers with practical requirements for achieving deep extinctions with PMCs. This is realized in two steps: firstly, an alternative mathematical development of complex amplitudes formed into the Lyot stop plane is described in section 2, and allows defining two necessary conditions for an azimuthal phase mask to achieve full extinction. Examples of applications are then given in section 3, including a review of various azimuthal phase functions (§ 3.1), hints for building more sophisticated functions (§ 3.2), and a simplified expression of the light leaks inside Lyot stop due to mask imperfection, suitable to system design and tolerancing (§ 3.3). A brief conclusion is drawn in section 4. The most laborious mathematical developments are given in an appendix.

## 2  MATHEMATICAL DEVELOPMENT

### 2.1  Basic hypotheses and coordinate systems

It is assumed in this study that the telescope pupil is illuminated with a monochromatic reference wavefront of wavenumber $k = 2\pi/\lambda$ with $\lambda$ the wavelength of the electromagnetic field. Therefore the conditions of full extinction are only valid for that single wavelength. Moreover the following hypotheses are made:

- The input wavefront is flat (i.e. originating from a point, unresolved source) and does not exhibit phase aberrations.

- It is spatially uniform over a circular pupil with no central obscuration, nor secondary mirror support legs.

- The phase mask is assumed to have infinite width.

- The analytical study is restricted to intensity distributions formed by the phase mask inside the Lyot stop plane. Other important performance metrics attached to the final coronagraph image plane, such as Inner working angles (IWA) or throughput (radial transmission curve of an off-axis planet) are not considered in the present paper.

For the sake of simplicity, the magnification factor between the input pupil and Lyot stop planes is taken equal to unity, and all focusing and collimating optics have an identical focal length denoted $F$. The employed coordinate systems are indicated in Figure 2 and consist in three reference frames **R**(OXYZ), **R'**(O'X'Y'Z) and **R''**(O''X''Y''Z), where Z is the optical axis of the system:

- **R** is attached to the exit pupil plane of the telescope OXY, with O the pupil centre. Points P in that plane are denoted either by their Cartesian coordinates $(x,y)$ or by their polar coordinates $(\rho,\theta)$.

- **R'** is attached to the telescope image plane O'X'Y' where the phase mask is located and centred at point O' along the optical axis. Points P' in the O'X'Y' plane are denoted by Cartesian coordinates $(x',y')$ or polar coordinates $(\rho',\theta')$.

- **R''** is attached to the Lyot stop plane O''X''Y'', with O'' the stop centre and points P'' are denoted by Cartesian coordinates $(x'',y'')$ or polar coordinates $(\rho'',\theta'')$.

### 2.2  Fourier optics analytic development

The theoretical frame of this study is restricted to first-order Gaussian optics and Fraunhofer scalar diffraction. In that case, the diffracted complex amplitude at any point M' located in the image plane O'X'Y' is proportional to the Fourier transform of the pupil transmission function, which writes in Cartesian coordinates:

$$A'(x',y') = K \int_{-\infty}^{+\infty}\int_{-\infty}^{+\infty} B_D(x,y)\exp\left[-2i\pi(xx' + yy')/\lambda F\right]dxdy, \qquad (1)$$

where K is a normalization factor and the pupil transmission function is a circular "pillbox" function of radius $R$ and diameter $D$ defined as:

$B_D(x,y) = 1$ $\quad$ when $0 \leq \rho = \sqrt{x^2 + y^2} \leq R = D/2$, and:

$B_D(x,y) = 0$ $\quad$ when $\rho > R$.

From now, we shall not use Cartesian coordinates any longer and express Eq. 1 in polar coordinates:

$$A'(\rho',\theta') = K \int_{0}^{2\pi}\int_{0}^{+\infty} B_D(\rho,\theta)\exp\left[-2i\pi\rho\rho'\cos(\theta'-\theta)/\lambda F\right]\rho d\rho d\theta. \qquad (2)$$

For a circular pupil the diffracted complex amplitude is axis symmetric and does not depend on $\theta$. Eq. 2 can thus be rewritten as:

$$A'(\rho') = K \int_0^{+\infty}\int_0^{2\pi} B_D(\rho) \exp[-2i\pi\rho\rho'\cos\theta/\lambda F] d\theta \rho d\rho = \pi R^2 \hat{B}_D(\rho'), \quad (3a)$$

where the function $\hat{B}_D(\rho')$ is defined as:

$$\hat{B}_D(\rho') = \frac{2J_1(2\pi R\rho'/\lambda F)}{2\pi R\rho'/\lambda F}, \quad (3b)$$

and $J_1$ is the type-J Bessel function at the first order. Classically the square modulus of $\hat{B}_D(\rho')$ is the well-known Airy function and the point-spread function (PSF) of the telescope. Inside the coronagraph, the complex amplitude $A''(\rho'',\theta'')$ formed at the Lyot stop plane O"X"Y" is obtained by multiplying $A'(\rho')$ with the complex transmission of the phase mask in the image plane, then taking the inverse Fourier transform of the result. Denoting $\phi'(\rho',\theta')$ the phase mask function and assuming infinite width, it yields:

$$A''(\rho'',\theta'') = K'\int_0^{2\pi}\int_0^{+\infty} \hat{B}_D(\rho')\exp[i\phi'(\rho',\theta')]\exp[2i\pi\rho'\rho''\cos(\theta''-\theta')/\lambda F]\rho' d\rho' d\theta'. \quad (4)$$

where K' is the normalization factor. Since our study is limited to the case of azimuthal phase mask coronagraphs, the phase mask function only depends on the angular coordinate $\theta$. Inserting Eq. 3b into Eq. 4, changing the integration variable $\rho'$ to $\nu$ with $\nu = \rho'/\lambda F$, and adjusting K' so that $A''(O'') = 1$ in the absence of phase mask (i.e. $\phi'(\theta') = 0$) finally gives an integral expression of the complex amplitude distribution in the Lyot stop plane:

$$A''(\rho'',\theta'') = R\int_0^{2\pi} \exp[i\phi'(\theta')] \int_0^{+\infty} J_1(2\pi R\nu)\exp[2i\pi\nu\rho''\cos(\theta''-\theta')] d\nu d\theta'. \quad (5)$$

Setting $\beta = \theta' - \theta''$ in Eq. 5 and defining the reduced coordinate $a = \rho''\cos\beta/R$ allows rewriting it in a more condensed form:

$$A''(\rho'',\theta'') = R\int_{-\theta''}^{2\pi-\theta''} \exp[i\phi'(\beta+\theta'')]F(a)d\beta, \quad (6a)$$

with function F(a) being defined as:

$$F(a) = \int_0^{+\infty} J_1(2\pi R\nu)\exp[2i\pi Ra\nu] d\nu. \quad (6b)$$

The geometrical interpretation of the parameter $a$ in Eqs. 6 is illustrated in Figure 3: at a given point P" of polar coordinates ($\rho''$,$\theta'$) in the Lyot stop plane, the complex amplitude $A''(\rho'',\theta'')$ is the azimuthal integral of the complex transmission of the phase mask weighted by function F(a). The condition $|a| = \rho''|\cos\beta|/R \le 1$ whatever the angle $\beta$ in a [0, 2π] interval defines a disk of radius R that corresponds to the telescope pupil. As demonstrated into the Appendix, the analytical expression of the weighting function F(a) is found to be:

$$F(a) = \frac{1}{2\pi R}\left(1 + i\frac{a}{\sqrt{1-a^2}}\right) \quad \text{for } |a| < 1, \quad \text{and:} \quad (7a)$$

$$F(a) = \frac{1}{2\pi R}\left(1 - \frac{|a|}{\sqrt{a^2-1}}\right) \quad \text{for } |a| > 1. \quad (7b)$$

Inserting Eqs. 7 into 6a gives an expression of the complex amplitude in the Lyot stop plane as function of the original variables ($\rho''$,$\theta'$):

$$A''(\rho'',\theta'') = \frac{1}{2\pi} \int_{\substack{\theta' \in [0,2\pi] \\ |a| \leq 1}} \left(1 + i\frac{\rho''\cos(\theta'-\theta'')}{\sqrt{R^2 - \rho''^2\cos^2(\theta'-\theta'')}}\right) \exp[i\phi'(\theta')]d\theta' \quad \text{for } \rho''|\cos(\theta'-\theta'')| \leq R, \quad (8a)$$

and:

$$A''(\rho'',\theta'') = \frac{1}{2\pi} \int_{\substack{\theta' \in [0,2\pi] \\ |a| > 1}}^{0} \left(1 - \frac{\rho''|\cos(\theta'-\theta'')|}{\sqrt{\rho''^2\cos^2(\theta'-\theta'') - R^2}}\right) \exp[i\phi'(\theta')]d\theta' \quad \text{for } \rho''|\cos(\theta'-\theta'')| > R. \quad (8b)$$

The previous equations are used in the two next sub-sections in order to evaluate the intensity distribution $I''(\rho'',\theta'') = |A''(\rho'',\theta'')|^2$ in the whole Lyot stop plane. They are illustrated geometrically in Figure 4 where the parameter $a = \rho''\cos\beta/R$ corresponds to the ratio $|O''H''|/R$. For any point P" in the Lyot stop plane, the point H" will run two full laps around the red dashed circles of the Figure when $\beta$ ranges from 0 to $2\pi$. Depending on the location of P" those circles may intersect the contour of the telescope pupil or not, hence two cases are distinguished:

- When point P" is located inside the image of the telescope pupil (Figure 4-1), the expression of the intensity $I''(\rho'',\theta'')$ only involves Eq. 8a and is established in § 2.3.

- When it is located outside of the pupil image (Figure 4-2) both Eqs.8 must be used. This case is treated in § 2.4.

Another illustration of the complex function $F(a)$ is given in image Figure 5, where its real and imaginary parts are plotted as function of the angle $\beta$ for different values of the $\rho''/R$ ratio.

## 2.3 Intensity distribution inside pupil image

For all points P" of polar coordinates $(\rho'',\theta'')$ located inside the image of the telescope pupil, the condition $|\rho''\cos\beta| \leq R$ holds and the complex amplitude in the Lyot stop plane is described by Eq. 8a. Using again the angular integration variable $\beta = \theta' - \theta''$ and defining the parameter $c = R/\rho''$ allows rewriting it as:

$$A''(\rho'',\theta'') = \frac{1}{2\pi} \int_{-\theta''}^{-\theta''+2\pi} \left(1 + i\frac{\cos\beta}{\sqrt{c^2 - \cos^2\beta}}\right) \exp[i\phi'(\beta+\theta'')]d\beta \quad (9)$$

Let us now make the two following assumptions:

1) Although initially defined over a [0,2π] interval, the phase mask function $\phi'(\theta')$ can be extended to the [–∞,+∞] range and made periodic with a period of 2π, eventually exhibiting discontinuities at $\theta' = 2n\pi$, where $n$ is any positive or negative integer. Then the integration bounds in Eq. 9 can be changed from [–$\theta'$,–$\theta'$+2π] to [–π,+π], which will be more convenient for the next calculations.

2) Then the complex mask transmission $\exp[i\phi'(\theta')]$ is also periodic and can be decomposed in Fourier series of coefficients $a_n$, with integer index $n$ spanning from –∞ to +∞. Hence it may be written as:

$$\exp[i\phi'(\theta')] = \sum_{n=-\infty}^{+\infty} a_n \exp[in\theta'] \quad (10a)$$

where by definition:
$$a_0 = \frac{1}{2\pi} \int_{-\pi}^{+\pi} \exp[i\phi'(\theta')]d\theta' \quad \text{is the mean value, and:} \tag{10b}$$

$$a_n = \frac{1}{2\pi} \int_{-\pi}^{+\pi} \exp[i\phi'(\theta')]\exp(-in\theta')d\theta' \quad \text{when } n \neq 0. \tag{10c}$$

Following both previous assumptions, and after inverting the continuous and discrete sums, the complex amplitude $A''(\rho'',\theta'')$ may be expressed as function of the coefficients $a_n$:

$$A''(\rho'',\theta'') = \frac{1}{2\pi} \sum_{n=-\infty}^{+\infty} a_n \int_{-\theta''}^{-\theta''+2\pi} \left(1 + i\frac{\cos\beta}{\sqrt{c^2 - \cos^2\beta}}\right) \exp[in(\beta + \theta'')]d\beta \tag{11}$$

or under a more condensed form:

$$A''(\rho'',\theta'') = \frac{1}{2\pi} \sum_{n=-\infty}^{+\infty} a_n (a_0 + ig_n(c)) \exp[in\theta''], \tag{12a}$$

with functions $g_n(c)$ being defined as:

$$g_n(c) = \int_{-\pi}^{+\pi} \frac{\cos\beta}{\sqrt{c^2 - \cos^2\beta}} \exp[in\beta]d\beta. \tag{12b}$$

From the geometrical interpretation of Figure 4-1, this integral is equal to twice its value over the integration range $[0,+\pi]$:

$$g_n(c) = 2\int_0^{+\pi} \frac{\cos\beta}{\sqrt{c^2 - \cos^2\beta}} \cos(n\beta)d\beta. \tag{13}$$

Setting the new variable $\theta = \beta - \pi/2$ in Eq. 13 then leads to:

$$g_n(c) = 2\cos(n\pi/2) \int_{-\pi/2}^{+\pi/2} \frac{\sin\theta}{\sqrt{c^2 - \sin^2\theta}} \cos(n\theta)d\theta - 2\sin(n\pi/2) \int_{-\pi/2}^{+\pi/2} \frac{\sin\theta}{\sqrt{c^2 - \sin^2\theta}} \sin(n\theta)d\theta. \tag{14}$$

Noting that the first term in Eq. 14 is null since the function to be integrated between $-\pi/2$ and $+\pi/2$ is odd, the final expression of $g_n(c)$ writes as:

$$g_n(c) = -i^n[1-(-1)^n]\int_{-\pi/2}^{+\pi/2} \frac{\sin\theta}{\sqrt{c^2 - \sin^2\theta}} \sin(n\theta)d\theta. \tag{15}$$

That last expression leads us to distinguish the cases when $n$ is either even or odd: Regrouping Eqs. 10a, 12 and 15 and after a few simplifications, the expressions of the complex amplitude into the telescope pupil plane finally are:

$$A''_{2p}(\rho'',\theta'') = \frac{a_0^2}{2\pi} \tag{16a}$$

when $n = 2p$ is an even integer, and:

$$A''_{2p+1}(\rho'',\theta'') = A''_{2p}(\rho'',\theta'') - \frac{i}{\pi} \sum_{p=-\infty}^{+\infty} (-1)^n a_n \exp[i(2p+1)\theta''] \int_{-\pi/2}^{+\pi/2} \frac{\sin\theta}{\sqrt{c^2 - \sin^2\theta}} \sin((2p+1)\theta)d\theta \quad (16b)$$

when $n = 2p+1$ is an odd integer. These expressions allow defining two sufficient conditions for an azimuthal phase function $\phi'(\theta')$ to generate full extinction inside the Lyot stop of a coronagraph. They are:

**C1**. The averaged complex amplitude $\exp[i\phi'(\theta')]$ is null over a periodic interval of $2\pi$ (i.e. $a_0 = 0$).

**C2**. The decomposition of $\exp[i\phi'(\theta')]$ into Fourier series shall only comprise even terms (all odd Fourier coefficient $a_{2p+1}$ are null).

These sufficient conditions call for the following comments:

1. **C1** and **C2** are necessary but not sufficient conditions, since it cannot be excluded that a given combination of odd coefficients $a_{2p+1}$ and functions $g_{2p+1}(c)$ produces a null amplitude $A''(\rho'',\theta'')$ in Eq. 12a.

2. The necessary condition **C1** alone only ensures null diffracted intensity at the Lyot stop centre O", as has been pointed out in Ref. [18].

3. Starting from a given phase function $\phi'(\theta')$ that satisfies **C1**, it shall be possible to build a family of $\pi/p$–periodic functions $\phi'(2p\theta')$ generating full extinction over the whole telescope pupil. This last property will be illustrated by some examples given in section 3.

It may finally be noted that when condition **C1** is not respected, the generated intensity inside the Lyot stop is uniform and deduced from Eq. 16a:

$$I''_{2p}(\rho'',\theta'') = |a_0^2|^2 / 4\pi^2 . \quad (17)$$

This light leak inside Lyot stop is further discussed in section 3.3.

### 2.4 Energy outside the Lyot stop

The case when point P" is located outside of the Lyot stop is a bit less interesting and is only treated here to check consistency with the results of Refs. [8] and [15]. The mathematical development is basically the same as in the previous sub-section, but here the expression of $g_n(c)$ in Eq. 13 is somewhat more complicated since the integration domain has to be divided into two different sections where either Eq. 8a or 8b shall be introduced into Eq. 6a, depending on the numerical value of the ratio $c = R/\rho''$ ($c \geq 1$ for points P" located inside the telescope pupil and $c < 1$ otherwise). Using the geometrical interpretation of Figure 4-2 for defining the integration bounds, noting that the integral sums are equal to twice their values over a single loop on the red circles and taking all symmetries into account, $g_n(c)$ becomes:

$$g_n(c) = 4 \int_0^{\arccos c} \frac{|\cos\beta|}{\sqrt{\cos^2\beta - c^2}} \cos(n\beta)d\beta + 2 \int_{\arccos c}^{\pi - \arccos c} \frac{\cos\beta}{\sqrt{c^2 - \cos^2\beta}} \cos(n\beta)d\beta . \quad (18)$$

Restricting us to the case when condition **C2** is respected, it can be shown that the second term of Eq. 18 is equal to zero, by use of the same mathematical arguments as in the previous sub-section. Moreover, only a VPM of topological charge $n = 2p$ is considered here. Then the Fourier decomposition of the function $\exp[i\phi'(\theta')]$ reduces to the vortex itself. Then applying condition **C1**, combining Eqs. 10c, 12 and 18, noting that $\cos\beta \geq 0$ on the considered integration domain and omitting the complex phase term in $n\theta'$, the expression of the complex amplitude outside of the telescope pupil simplifies as:

$$A''_n(\rho'',\theta'') = \frac{2}{\pi} \int_0^{\arccos c} \frac{\cos\beta}{\sqrt{\cos^2\beta - c^2}} \cos(n\beta) d\beta. \quad (19)$$

The latter can further be reduced to a polynomial expansion of the term $c^2$ by use of three successive tricks:

1) Since $n = 2p$ is an even integer, develop $\cos(n\beta)$ into even powers of $\cos\beta$ following the general relation $\cos(n\beta) = \sum_{k=0}^{p} b_k \cos^{2k}\beta$ where the coefficients $b_k$ are known real numbers.

2) Inside the integrals, set the new variables $u = \sin\beta / \sqrt{1-c^2}$, then $\theta = \arcsin u$ leading to a polynomial expression of the terms $c^{2k}$:

$$A''_n(\rho'',\theta'') = \frac{2}{\pi} \sum_{k=0}^{p} b_k \int_0^{\pi/2} \left[1 - (1-c^2)\sin^2\theta\right]^k d\theta. \quad (20)$$

3) Finally use the recurrence relationships $S_n = \int_0^{\pi/2} \sin^n\theta\, d\theta = \frac{n-1}{n} S_{n-2}$ and $S_0 = \pi/2$ to evaluate the integrals after binomial development of $\left[1 - (1-c^2)\sin^2\theta\right]^k$ in Eq. 20.

The previous procedure has been applied to a VPM of topological charges $m = 2$, 4 and 6 for validation purpose. Expressions of the amplitudes $A''_m(\rho'',\theta'')$ and of their main parameters are compiled in Table 1. Although obtained in a much less elegant manner, these expressions are rigorously similar to those given into the Appendix B of Ref. [15]. Plots of the amplitudes $A''_n(\rho'',\theta'')$ and resulting intensities $I''_n(\rho'',\theta'') = |A''_n(\rho'',\theta'')|^2$ outside of the telescope pupil are reproduced in Figure 6.

Table 1: Expressions of amplitudes radiated outside of the telescope pupil for a VPM of topological charges $m = 2$, 4 and 6, as function of parameter $c = R/\rho''$.

| Vortex topological charge $m = 2p$ | $p = m/2$ | $c_k$ coefficients ($0 \leq k \leq p$) | $S_{2k}$ integrals ($0 \leq k \leq p$) | $A''_n(\rho'',\theta'')$ |
|---|---|---|---|---|
| 2 | 1 | −1 and +2 | $\pi/2$ and $\pi/4$ | $c^2$ |
| 4 | 2 | +1, −8 and +8 | $\pi/2$, $\pi/4$ and $3\pi/16$ | $-2c^2 + 3c^4$ |
| 6 | 3 | −1, +8, −48 and +32 | $\pi/2$, $\pi/4$, $3\pi/16$ and $5\pi/32$ | $3c^2 - 12c^4 + 10c^6$ |

## 3 EXAMPLES OF APPLICATIONS

In this section are firstly recalled some elementary phase masks functions employed in coronagraphy, such as vortex, azimuthal cosine, and regularly segmented functions (§ 3.1). The way those functions can be combined together is briefly discussed in § 3.2. Finally, an empirical formula relating phase mask errors to resultant light leaks inside the Lyot stop is defined in § 3.3.

### 3.1 Elementary phase masks functions

*3.1.1 Vortex functions*

Nowadays the most popular azimuthal phase function is probably the vortex function that was firstly introduced in the field of coronagraphy by Swartzlander [7] and is also commonly used for particles trapping [19]. It is a very simple linear phase ramp written as:

$$\phi'(\theta') = m\theta', \qquad (21)$$

where $m$ is a positive or negative integer named topological charge. It follows that the Fourier decomposition of the associated complex amplitude $\exp[i\phi'(\theta')]$ reduces to the sole non-zero coefficient $a_m = 1$. Therefore, vortex complex amplitudes can be considered as the basis for all other amplitude distributions generating extinctions inside the Lyot stop of the coronagraph, as pointed out in Refs. [8] and [15]. Obviously, this statement is not applicable to the $\phi'(\theta')$ phase function itself.

The major properties of optical vortex PMCs of topological charges $m = 1, 2, 3$ and $4$ are illustrated in Figure 7-1 to 7-4 respectively. For each case, three-dimensional plots of the phase functions $\phi'(\theta')$ are displayed on the left column. The functions are scaled in terms of waves, i.e. $\phi'(\theta')/2\pi$. The modules of the Fourier coefficients $a_n$ associated to the complex amplitude $\exp[i\phi'(\theta')]$ are illustrated by the histograms of the central column. They have been computed numerically by direct application of Eqs. 10. Finally, the right column of the figure exhibits grey-scale maps of the intensity distribution $I''(\rho'', \theta'')$ generated in the Lyot stop plane. They have been computed using a double Matrix Fourier transform (MFT) algorithm described in Ref. [20] with input pupil sampling of 4096 x 4096 embedded into 8192 x 8192 arrays to prevent aliasing effects. The intensity maps are shown in logarithmic scale for better visualization of the central extinction area. They reveal the clear superiority of vortices of even topological charges ($m = 2$ and $4$) with respect to odd charges ($m = 1$ and $3$) for achieving deep extinction into the telescope pupil area. These results are in agreement with the necessary conditions **C1** and **C2** of sub-section 2.3, because even phase functions respect both of them, while odd functions respect condition **C1** only. It can be noticed however that central extinction tends to improve with increasing vortex charges, thus it might be conjectured that odd charges $m = 2p + 1$ could also be suitable to coronagraphs for increasing odd vortex charges.

*3.1.2 Azimuthal cosine-modulated functions*

The azimuthal cosine-modulated (ACM) phase function is already known for possible applications in laser beam shaping and particles manipulation [21-22] and has recently been proposed for coronagraphic applications [18] [23]. The main difference with respect to vortex or regularly segmented coronagraphs is that they exhibit a single indeterminate point at O' instead of one or several discontinuity lines. Mathematically, the ACM function is defined as:

$$\phi'(\theta') = z_m \cos(k\theta'), \qquad (22)$$

where $k$ is an integer number named as the function frequency, and $z_m$ is the $m^{th}$ zero of the type-J Bessel function $J_0(z)$. The general properties of the ACM functions are illustrated in Figure 8. As for vortex case, the histograms in the central column of the figure show the modules of the Fourier coefficients $a_n$ when $k = 1, 2, 3$ and $4$. Their expression can be deduced from the case $k = 1$ by inserting $\phi'(\theta')$ into Eq. 10c and using the integral definition of type-J Bessel functions, finally leading to $a_n = (-i)^n J_n(z_m)$. Therefore the necessary condition **C1** is fulfilled since the coefficient $a_0$ is always equal to zero, while condition **C2** is respected for all even integers $k$. It can be verified in Figure 8 that the $a_n$ coefficients are identical to their theoretical values, and that full extinction inside the Lyot stop is only achieved when $k = 2$ and $4$. As illustrated by their radial and azimuthal profiles shown in Figure 9, the intensity distributions produced in the Lyot stop plane are no longer axis symmetric, due to the presence of different non null Fourier coefficients. More details about the ACM phase masks will be found in the section 3.2 of Ref. [18], where they have been compared to the VPM in terms of achievable extinction rate, inner working angle and throughput. It was concluded that despite of slightly degraded performance, ACM functions represent an interesting alternative to the VPM, especially because they may be easier to produce as phase plates or by using deformable mirrors.

*3.1.3 Regularly segmented functions*

Historically, the first azimuthal phase functions proposed for coronagraphy were regularly segmented functions defined as:

$$\phi'(\theta') = \pi \, \text{Int}(k\theta'/\pi), \qquad (23)$$

where function Int($z$) stands for the integer part of any real number $z$, and $k$ is an integer equal to half the number of segments. In particular, the phase knife [24], four-quadrant [6] and eight-octant [25] coronagraphs are obtained when $k$ = 1; 2 and 4 respectively. They are all illustrated in Figure 10 and Figure 11, together with the less utilized six-sextant ($k$ = 3). Once again, it can be verified that full extinction is only achieved for even $k$ numbers. Here the theoretical expression of the Fourier coefficients $a_n$ is not particularly simple, but its compliance with the computed histograms has also been checked.

## 3.2 Composite phase functions

All phase functions presented in the last sub-sections may be considered as basic bricks for building more elaborated functions, also generating full extinction inside the Lyot stop of the coronagraph, such as the composite phase function described in Ref. [26]. The way these functions can be combined for generating central extinctions (i.e. fulfilling condition **C1** only) has already been described in Ref. [22]. In particular it was demonstrated that:

   a. Central nulls generated by azimuthal phase functions are not affected by axis-symmetric central obscurations.
   b. In case of angular truncation (i.e. $\theta$ ranging from arbitrary segments [$\theta_1, \theta_2$] instead of [0,2π]), central nulls can be preserved by proper rescaling of the phase function amplitude.
   c. Different types of such phase functions can be juxtaposed on different angular segments and still generate a central null.

For phase mask coronagraphs the introduction of additional condition **C2** implies severe limitations and few of the above properties shall remain valid. For example a central obscuration of radius $\rho'_0$ shall turn the integration domain [0,+∞] in Eq. 6b into [$\rho'_0$,+∞], thus making the expressions of function $F(a)$ in Eqs. 7 invalid, and so property a. Similarly, restricting the integration domain [0,2π] in relation 12b will modify the expressions of $A''(\rho'', \theta'')$ in Eqs. 16, which contradicts property b. Hence the phase masks should only be non obstructed disks. It remains possible, however, to define composite phase functions built from the three different types identified in section 3.1 and achieving full extinction into the Lyot stop. Figure 12 illustrates three examples of such functions that are analytically defined in Table 2. The general procedure for constructing them consists in firstly identifying a function $\phi'(\theta')$ satisfying condition **C1**, then building the function $\phi'(2p\theta')$ where $p$ is an integer as explained in § 2.3. In that way there certainly exists an infinity of composite nulling phase functions $\phi'(\theta')$ suitable to coronagraphy, albeit exhibiting different intensity distributions outside of the Lyot stop as shown in the rightmost column of the figure.

Table 2: Three examples of composite nulling phase functions $\phi'(\theta')$ illustrated in Figure 12.

| Number | Analytic expression | | Remark |
|---|---|---|---|
| 1 | $\phi'(\theta') = 4\theta' + \pi \, \text{Int}(4\theta'/\pi)$ | | Addition of a charge 4 vortex with a four-quadrant |
| 2 | $\phi'(\theta') = 4\theta'$ | –π ≤ $\theta'$ < –π/2 and 0 ≤ $\theta'$ < +π/2 | Composition of vortex and ACM functions over four π/2 segments |
|   | $\phi'(\theta') = z_1 \cos(4\theta')$ | –π/2 ≤ $\theta'$ < 0 and +π/2 ≤ $\theta'$ < +π | |
| 3 | $\phi'(\theta') = \pi \, \text{Int}(8\theta'/\pi)$ | –π ≤ $\theta'$ < –π/2 and 0 ≤ $\theta'$ < +π/2 | Composite eighth-octant phase function |
|   | $\phi'(\theta') = \pi \, \text{Int}(8(\theta' - \pi/2)/\pi)$ | –π/2 ≤ $\theta'$ < 0 and +π/2 ≤ $\theta'$ < +π | |

## 3.3 Light leaks inside Lyot stop

Although the analytical formalism of this study is restricted to the case of an aberration-free telescope, i.e. with no phase errors introduced into the pupil plane OXY nor in any intermediate plane located between points O and O' (usually requiring Fresnel diffraction analysis), it turns out that it is usable for introducing phase defects into the image plane O'X'Y'. Such aberrations may for instance result from phase mask manufacturing errors and are usually neglected or treated numerically. In that case however an approximate formula can be established for evaluating the effect of random phase mask errors on the achievable extinction depth inside the Lyot stop. For phase functions $\phi'(\theta')$ satisfying both conditions **C1** and **C2**, it can be written from Eq. 17:

$$\langle I''_{2p}(\rho'', \theta'') \rangle = \langle |a_0^2|^2 \rangle / 4\pi^2 , \tag{24a}$$

where the symbol $\langle \ \rangle$ stands for statistical average, and phase errors $\delta\phi'(\theta')$ are introduced into the expression of $a_0$:

$$a_0 = \frac{1}{2\pi} \int_{-\pi}^{+\pi} \exp[i(\phi'(\theta') + \delta\phi'(\theta'))] d\theta' \tag{24b}$$

Assuming weak phase distortions, i.e. $\delta\phi'(\theta')/\phi'(\theta') \ll 1$ the complex exponential $\exp[i\,z]$ can be approximated to $1 + i\,z$, thus leading to:

$$a_0 \approx \frac{i}{2\pi} \int_{-\pi}^{+\pi} \delta\phi'(\theta') \exp[i\phi'(\theta')] d\theta' . \tag{25}$$

Applying Schwarz's inequality allows obtaining a worst-case estimation of $|a_0|$:

$$|a_0| \approx \left| \int_{-\pi}^{+\pi} \delta\phi'(\theta') \exp[i\phi'(\theta')] d\theta' \right| \leq \int_{-\pi}^{+\pi} |\delta\phi'(\theta')| d\theta' \approx \sigma , \tag{26}$$

where $\sigma$ is the standard deviation of the phase mask errors assumed to be Gaussian. It must be noted that the pessimistic use of Schwarz's inequality is somewhat counterbalanced by the optimistic assumption of neglecting energy leaks carried by the other Fourier coefficients $a_n$. An approximate expression for the averaged intensity distribution inside the Lyot stop is then:

$$\langle I''_{2p}(\rho'', \theta'') \rangle \approx \sigma^4 / 4\pi^2 . \tag{27}$$

In Figure 13 are illustrated the effects of random phase errors added to a vortex phase mask of charge $m = 2$, both on its Fourier coefficients $a_n$ and intensity distribution generated in the Lyot stop plane. The plots are shown in logarithmic scale for better enhancing low values. Figure 13-3 and 4 show plots of $\langle I''_{2p}(\rho'', \theta'') \rangle$ computed from Eq. 27 in red solid lines, compared with the results of numerical simulations indicated with red dots. Curves are plotted as function of the standard deviation $\sigma$ in linear and logarithmic scale respectively. It can be noted that simulation results are in global agreement with the approximate formula 27, even if data do not match perfectly. The remaining deviations most probably originate from the three employed approximations (weak phase errors, use of Schwarz's inequality, and Fourier decomposition reduced to the single coefficient $a_0$), that may also explain non-uniform irradiance inside the Lyot stop shown in Figure 13-2. Despite of the limitations of the analytical model, it can be concluded however that:

- Resulting from the fourth-power law in relation 27, light leaks are increasing very rapidly with phase errors .

- Assuming necessary extinction ratios of $10^{-6}$ and $10^{-10}$ for observing Jupiter or Earth-like extra-solar planets respectively, phase plates manufacturing requirements should be 7.9 $10^{-2}$ and 7.9 $10^{-3}$ rad, or $\lambda/79$ and $\lambda/793$ equivalently (all numbers are in RMS terms).

- Knowing that phase masks employed in coronagraphy usually have small areas (typically 100 mm$^2$), these specifications look difficult but feasible for Jupiter-like extra-solar planets. The case of Earth-like planets still remains very challenging.

At this stage it may also be of interest to evaluate the effects of local, deterministic phase mask errors on the intensity distributions formed in the final image plane of the coronagraph. Since Eq. 27 is no longer relevant, the question is addressed numerically with the help of the numerical model described in § 3.1.1: starting from the complex amplitude $A''(\rho'', \theta'')$ in the Lyot stop plane, one third and last Fourier transform allows computing irradiance maps in the coronagraph image plane. The results of this simulation are illustrated in Figure 14. Local phase mask errors of amplitude $\delta\phi' \approx \pm 0.4\lambda$ were introduced at random locations into the phase mask plane O'X'Y'. The resulting intensity distribution in the final image plane (referred to the X' and Y' axes, assuming a magnification factor of one between both image planes) are shown in Figure 14-2. In addition to an expected star leakage on the optical axis, it can be seen that there is a correlation between local errors of the phase mask and residual speckles in the coronagraph image plane, on the one hand, and that the strength of these speckles is inversely proportional to their distance to point O', on the other hand. Such parasitic off-axis images of the star could eventually be mistaken with speckles generated by quasi-static pupil aberrations. The way of disentangling those two different types of static speckles is however beyond the scope of the paper.

## 4 CONCLUSION

In this paper was presented an analytical study of the azimuthal phase-mask coronagraph, one of the most promising instrumental concepts envisioned for detecting and characterizing Jupiter or Earth-like extra-solar planets in the near future. Special emphasis was put on the physical and geometrical interpretation of the mathematical development. It allowed defining two necessary conditions for the phase mask to achieve full extinction in the pupil plane of the coronagraph. They state that the complex amplitude generated by the azimuthal phase mask should have zero average, on the one hand, and that its Fourier coefficients should only be even, on the other hand. Examples of such phase functions have been reviewed, including optical vortices, four-quadrant phase mask, and the less utilized azimuthal cosine-modulated functions. Some hints for building more sophisticated functions were given, suggesting that there exist huge varieties of phase functions satisfying to the necessary conditions. Finally, a simplified expression of the light leaks due to mask imperfection and suitable to system design and tolerancing was proposed. It highlights the extreme difficulty of manufacturing phase masks suitable to the detection of Earth-like extra-solar planets.

The author would like to thank his colleague P. Rabou for careful reading of the manuscript and improvement of English wording. He also thanks the anonymous reviewers for relevant comments leading to considerable improvements of the original manuscript.

## REFERENCES

[1] J.-L. Beuzit, M. Feldt, K. Dohlen, D. Mouillet *et al*, "SPHERE: a planet finder instrument for the VLT," Proceedings of the SPIE vol. 7014, n° 701418 (2008).

[2] B. Macintosh, J. Graham, D. Palmer *et al*, "The Gemini Planet Imager," Proceedings of the SPIE vol. 6272, n° 62720L (2006).

[3] F. Zhao, "WFIRST-AFTA coronagraph instrument overview," Proceedings of the SPIE vol. 9143, n° 91430O (2014).

[4] F. Roddier, C. Roddier, "Stellar coronagraph with phase mask," Publications of the Astronomical Society of the Pacific vol. 109, p. 815-820 (1997).

[5] C. Aime, R. Soummer, A. Ferrari, "Total coronagraphic extinction of rectangular apertures using linear prolate apodizations," Astronomy and Astrophysics vol. 389, p. 334-344 (2002).

[6] D. Rouan, P. Riaud, A. Boccaletti, Y. Clénet, A. Labeyrie, "The four-quadrant phase-mask coronagraph. I. Principle," Publications of the Astronomical Society of the Pacific vol. 112, p. 1479-1486 (2000).
[7] G. A. Swartzlander Jr, "Peering into darkness with a vortex spatial filter," Optics Letters vol. 26, p. 497-499 (2001).
[8] D. Mawet, P. Riaud, O. Absil, J. Surdej, "Annular groove phase mask coronagraph," Astrophysical Journal vol. 633, p. 1191-1200 (2005).
[9] G. A. Swartzlander Jr, "The optical vortex coronagraph," J. Opt. A: Pure Appl. Opt. vol. 11, 094022 p. 1-9 (2009).
[10] F. Lemarquis, M. Lequime, G. Albrand, L. Escoubas, J.-J. Simon, J. Baudrand, P. Riaud, D. Rouan, A. Boccaletti, P. Baudoz, D. Mawet, "Manufacturing of 4-quadrant phase mask for nulling interferometry in thermal infrared," Proceedings of the SPIE vol. 5250, p. 435-443 (2004).
[11] D. Mawet, E. Serabyn, K. Liewer, C. Hanot, S. McEldowney, D. Shemo, N. O'Brien, "Optical vectorial vortex coronagraphs using liquid crystal polymers: theory, manufacturing and laboratory demonstration," Optics Express vol. 17 p. 1902-1918 (2009).
[12] A. Boccaletti, P. Riaud, P. Baudoz, J. Baudrand; D. Rouan, D. Gratadour, F. Lacombe, A.-M. Lagrange, "The four-quadrant phase mask coronagraph. IV. First light at the Very Large Telescope," Publications of the Astronomical Society of the Pacific vol. 116, p. 1061-1071 (2004).
[13] D. Mawet, E. Serabyn, K. Liewer, R. Burruss, J. Hickey, D. Shemo, "The vector vortex coronagraph: Llaboratory results and first light at Palomar Observatory," Astrophysical Journal vol. 709,:p. 53-57 (2010).
[14] C. Jenkins, "Optical vortex coronagraphs on ground-based telescopes," Mon. Not. R. Astron. Soc. 384, 515–524 (2008).
[15] A. Carlotti, G. Ricort, C. Aime, "Phase mask coronagraphy using a Mach-Zehnder interferometer," Astronomy and Astrophysics vol. 504, p. 663-671 (2009).
[16] I. M. Gel'fand, G. E. Shilov, *Generalized functions*, Academic Press (1966–1968).
[17] M. Abramowitz, I. A. Stegun, *Handbook of mathematical functions*, Dover Publications, INC., New York (1972).
[18] F. Hénault, A. Carlotti, C. Vérinaud, "Analysis of nulling phase functions suitable to image plane coronagraphy," Proceedings of the SPIE vol. 9912, n° 99126K (2016).
[19] 1. A. M. Yao, M. J. Padgett, "Orbital angular momentum: origins, behavior and applications," Advances in Optics and Photonics vol. 3, p. 161-204 (2011).
[20] R. Soummer, L. Pueyo, A. Sivaramakrishnan, R. J. Vanderbei, "Fast computation of Lyot-style coronagraph propagation," Optics Express vol. 15, p. 15935-15951 (2007).
[21] S. Topuzoski, L. Janicijevic, "Diffraction characteristics of optical elements designed as phase layers with cosine-profiled periodicity in the azimuthal direction," JOSA A vol. 28, p. 2465-2472 (2011).
[22] F. Hénault, "Strehl ratio: a tool for optimizing optical nulls and singularities," JOSA A vol. 32, p. 1276-1287 (2015).
[23] O. Ma, Q. Cao, F. Hou, "Wide-band coronagraph with sinusoidal phase in the angular direction," Optics Express vol. 20, p. 10933-10943 (2012).
[24] L. Abe, F. Vakili1, A. Boccaletti, "The achromatic phase knife coronagraph," Astronomy and Astrophysics vol. 374, p. 1161-1168 (2001).
[25] N. Murakami, J. Nishikawa, K. Yokochi, M. Tamura, N. Baba, L. Abe, Achromatic eight-octant phase-mask coronagraph using photonic crystal," The Astrophysical Journal vol. 714, p. 772-777 (2010).
[26] F. Hou, Q. Cao, M. Zhu, O. Ma, "Wide-band six-region phase mask coronagraph," Optics Express vol. 22, p. 1884-1895 (2014).

# APPENDIX. CALCULATION OF FUNCTION $F(a)$

Let us first make use of the integral expression of the $J_1$ Bessel function that is:

$$J_1(z) = -\frac{i}{2\pi} \int_0^{2\pi} \exp[i(\theta + z\cos\theta)] d\theta. \tag{A1}$$

Inserting Eq. A1 into 6b and permuting the summation signs readily leads to:

$$F(a) = -\frac{i}{2\pi} \int_0^{2\pi} \exp[i\theta] \hat{H}(a,\theta) d\theta, \tag{A2}$$

where:
$$\hat{H}(a,\theta) = \int_0^{+\infty} \exp[2i\pi R\nu(a+\cos\theta)] d\nu. \tag{A3}$$

The function $\hat{H}(a,\theta)$ in Eq. A3 actually appears as the Fourier transform of the Heaviside distribution $H(z)$ at $z = -R(a+\cos\theta)$. Knowing that the Fourier transform of $H(z)$ is equal to $\delta(z)/2 - i/2\pi z$, where $\delta(z)$ is the impulse Dirac distribution, $F(a)$ can be rewritten as:

$$F(a) = -\frac{i}{4\pi} \int_0^{2\pi} \delta(-R(a+\cos\theta)) \exp[i\theta] d\theta + \frac{1}{4\pi^2 R} \int_0^{2\pi} \frac{\exp[i\theta]}{a+\cos\theta} d\theta$$
$$= -\frac{i}{4\pi} F_1(a) + \frac{1}{4\pi^2 R} F_2(a) \tag{A4}$$

with:
$$F_1(a) = \int_0^{2\pi} \delta(-R(a+\cos\theta)) \exp[i\theta] d\theta \tag{A5a}$$

and:
$$F_2(a) = \int_0^{2\pi} \frac{\exp[i\theta]}{a+\cos\theta} d\theta. \tag{A5b}$$

The analytical expressions of $F_1(a)$ and $F_2(a)$ are then evaluated separately.

**Analytical expression of $F_1(a)$**

Let us firstly rewrite $F_1(a)$ as:

$$F_1(a) = \text{Re}[F_1(a)] + \text{Im}[F_1(a)] = \int_0^{2\pi} \delta(-R(a+\cos\theta))\cos\theta \, d\theta + i \int_0^{2\pi} \delta(-R(a+\cos\theta))\sin\theta \, d\theta \tag{A6}$$

with Re[ ] and Im[ ] respectively denoting the real and imaginary parts of a complex number. We shall then make use of the general relationship [16]:
$$\delta(f(\theta)) = \sum_{n=1}^{N} \frac{\delta(\theta - \theta_n)}{|f'(\theta_n)|}, \tag{A7}$$

where $f'(\theta)$ is any real function of the variable $\theta$, $\theta_n$ with $1 \leq n \leq N$ its roots over the integration range, and $f''(\theta)$ the derivative of $f'(\theta)$. For $f(\theta) = -R(a + \cos\theta)$ the number of roots is depends on the value of $|a|$ and is either equal to 0 or 2:

- when $|a| \leq 1$, $N = 2$, $\theta_1 = \pi - \arccos a$ and $\theta_2 = \pi + \arccos a$.,
- when $|a| > 1$, $N = 0$ and both functions $\delta(-R(a + \cos\theta))$ and $F_1(a)$ are null.

When $|a| \leq 1$, combining Eqs A6 and A7 and using the elementary relation $\sin(\arccos a) = \sqrt{1 - a^2}$ finally leads to:

$$\text{Re}[F_1(a)] = -\frac{2a}{R\sqrt{1-a^2}} \qquad \text{and:} \qquad \text{Im}[F_1(a)] = 0 \tag{A8}$$

**Analytical expression of $F_2(a)$**

Here also $F_2(a)$ may be rewritten as:

$$F_2(a) = \text{Re}[F_2(a)] + i\,\text{Im}[F_2(a)] = \int_0^{2\pi} \frac{\cos\theta}{a + \cos\theta}\,d\theta + i\int_0^{2\pi} \frac{\sin\theta}{a + \cos\theta}\,d\theta. \tag{A9}$$

Changing the variable $t = \cos\theta$ in $\text{Im}[F_2(a)]$ readily shows that it is equal to zero, therefore the expression of $F_2(a)$ reduces to its real part that is, after elementary arithmetic and changing integration bounds:

$$F_2(a) = \text{Re}[F_2(a)] = 2\pi - a\int_{-\pi}^{+\pi} \frac{d\theta}{a - \cos\theta}. \tag{A10}$$

Then from Ref. [17], section 4.3.133, the expressions of $F_2(a)$ are:

when $|a| \leq 1$: $\quad F_2(a) = 2\pi - \frac{a}{\sqrt{1-a^2}}\left[\log\left(\frac{(1+a)\tan\theta/2 - \sqrt{1-a^2}}{(1+a)\tan\theta/2 + \sqrt{1-a^2}}\right)\right]_{-\pi}^{+\pi} = 2\pi$, (A11a)

when $|a| > 1$: $\quad F_2(a) = 2\pi - \frac{2a}{\sqrt{a^2-1}}\left[\arctan\left(\frac{(a+1)\tan\theta/2}{\sqrt{a^2-1}}\right)\right]_{-\pi}^{+\pi} = 2\pi\left(1 - \frac{|a|}{\sqrt{a^2-1}}\right)$. (A11b)

**Final expression of $F(a)$**

Combining Eqs. A4, A6, A8 and A11 finally leads to the analytical expression of $F(a)$ that is:

$$F(a) = \frac{1}{2\pi R}\left(1 + i\frac{a}{\sqrt{1-a^2}}\right) \qquad \text{when } |a| < 1, \quad \text{and:} \tag{A12a}$$

$$F(a) = \frac{1}{2\pi R}\left(1 - \frac{|a|}{\sqrt{a^2-1}}\right) \qquad \text{when } |a| > 1. \tag{A12b}$$

It should be noted that similar expressions can be derived from Ref. [17], sections 11.4.37 and 11.4.38, but their validity ranges are restricted to $0 < a < 1$ and $a > 1$, respectively for Eqs. A12a and A12b.

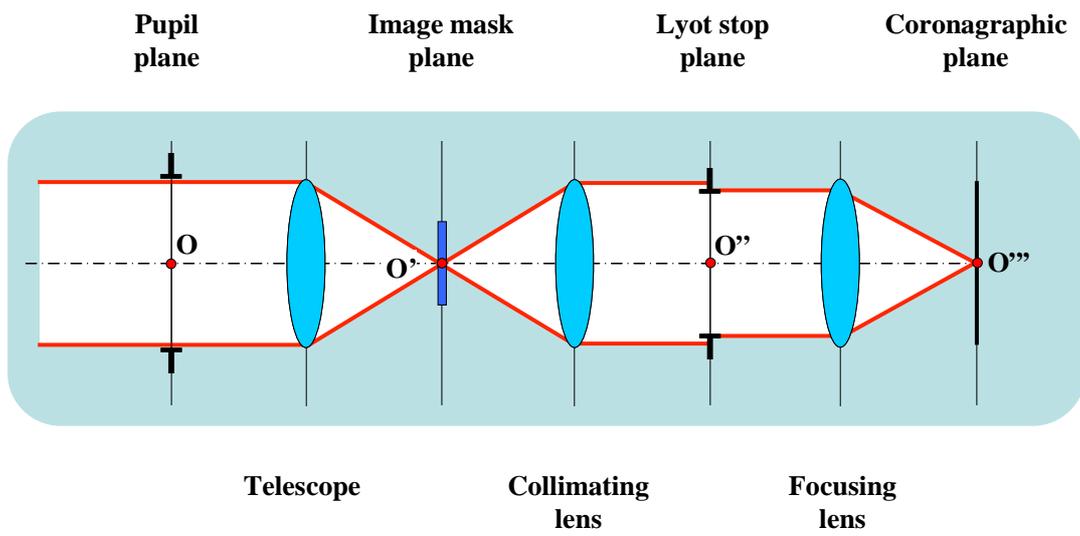

**Figure 1: Basic principle of a phase-mask coronagraph.**

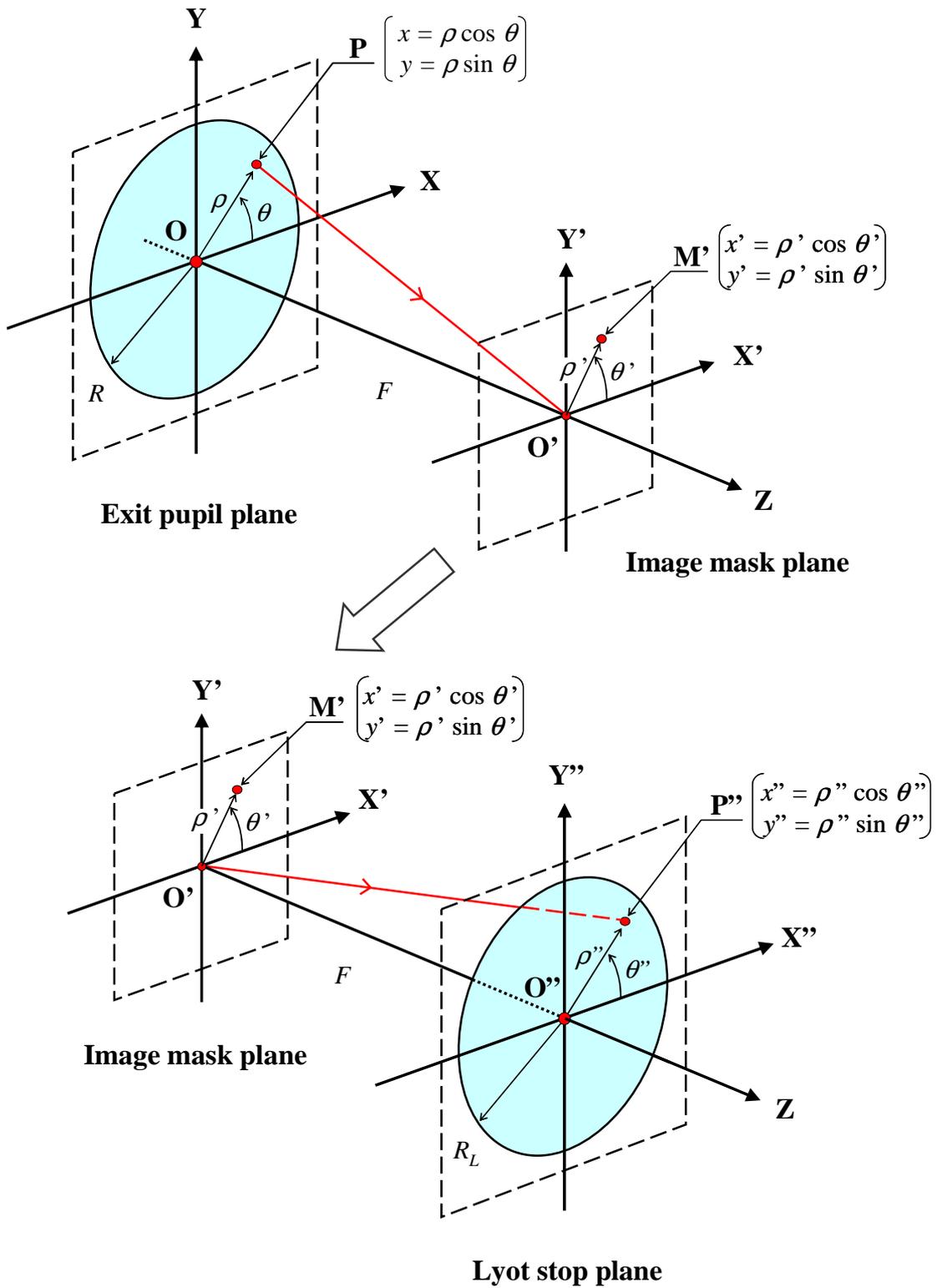

**Figure 2: Coordinate frames and scientific notations.**

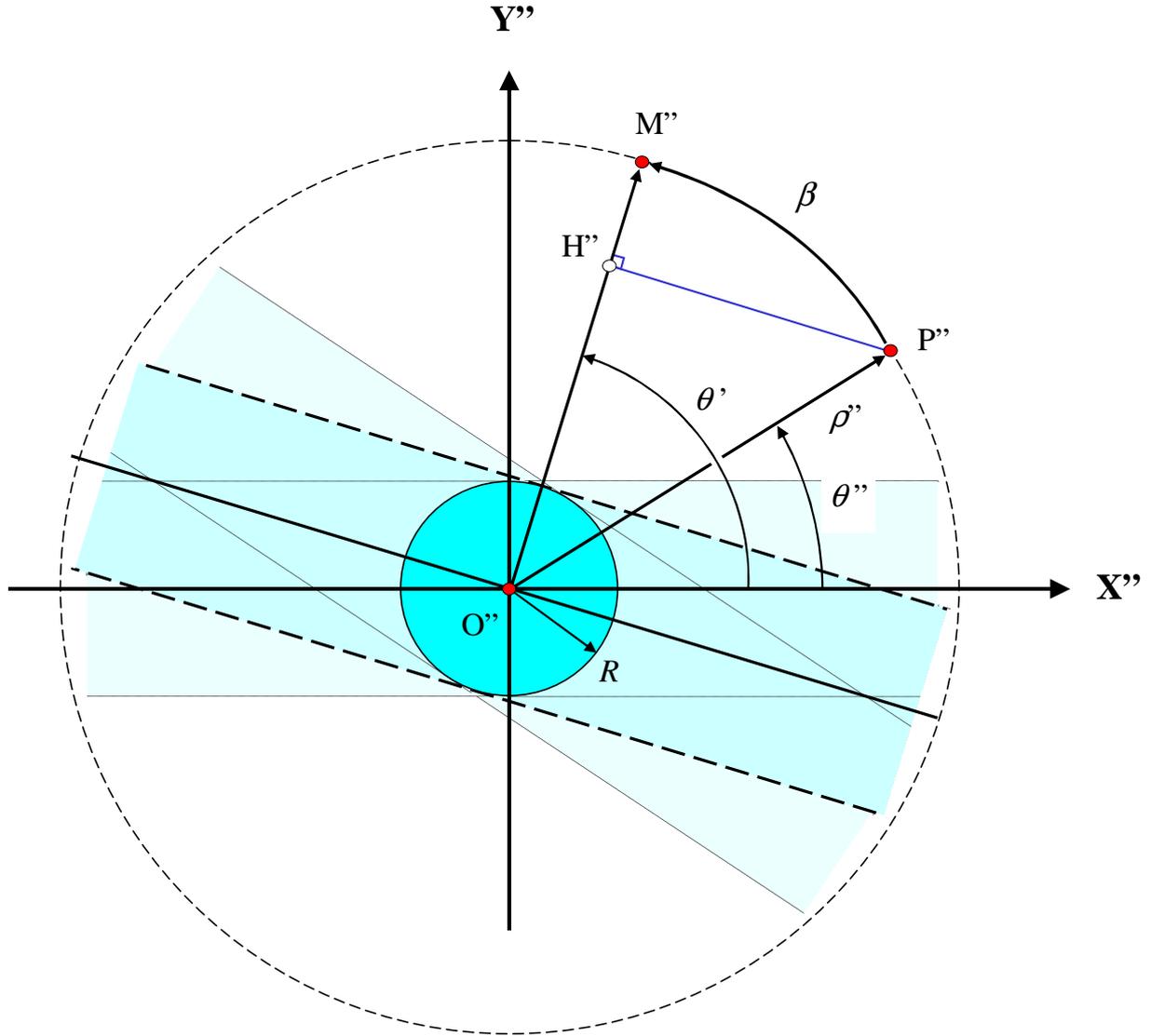

**Figure 3:** Geometrical interpretation of the parameter $a = \rho'' \cos\beta / R$. For a given point P" of polar coordinates $(\rho'', \theta'')$ in the Lyot stop plane O"X"Y", the integration domain of $A''(\rho'', \theta'')$ is defined as $0 \le \theta' \le 2\pi$ or $-\theta'' \le \beta \le 2\pi - \theta''$ (Eq. 6a). It can be represented as a point M" rotating by the angle $\theta''+\beta$ along a dashed-line circle of radius $\rho''$. The quantity $a = \rho'' \cos\beta / R$ is equal to the ratio $|O''H''|/R$, with H" the projection of P" on segment O"M". At a given angle $\beta$ the inequality $|a| \le 1$ refers to an area of infinite length and width $2R$ (shown in light blue). As $\beta$ ranges from 0 to $2\pi$ those areas intersect at the blue central disk of radius $R$, which is the geometrical image of the telescope pupil. Similarly, the condition $|a| > 1$ applies to points P" located outside of the pupil.

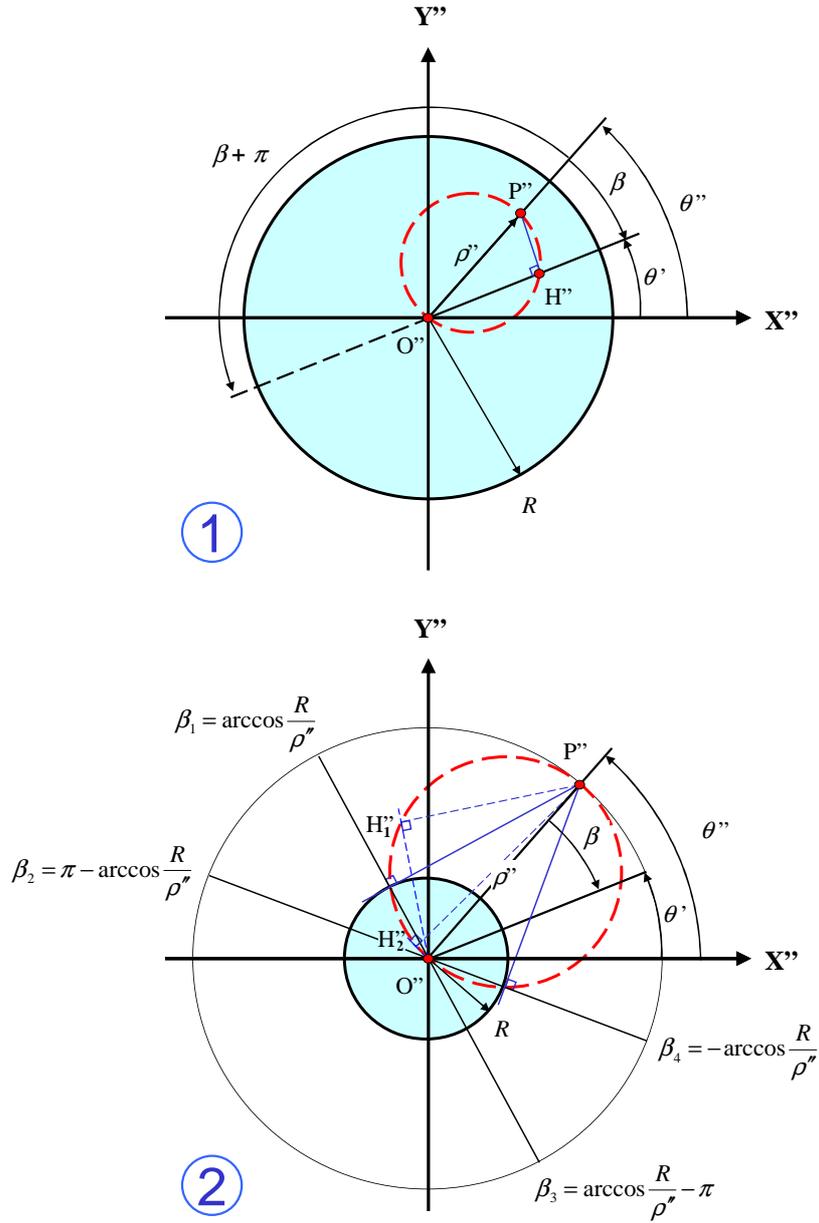

**Figure 4:** Definition of $\beta$ integration domains for points P" located inside the telescope pupil (1) and outside of it (2). For a given point P" of polar coordinates ($\rho$',$\theta$'), points H" move along the red dashed circles, running two laps when $\beta$ ranges from 0 to $2\pi$. <u>Case 1</u>: H" always remains within the image of the telescope pupil and the condition $|a| \leq 1$ is respected, then only Eq. 8a is needed for computing $I''(\rho'',\theta'')$. Moreover the integration range can be restricted to $[0,\pi]$ which stands for a single loop on the red circle (same points H" are found for any couple of angles $\beta$ and $\beta + \pi$). <u>Case 2</u>: the red circle intersects the contour of the telescope pupil. The $[0,2\pi]$ integration range must be divided into four sub-domains indicated by the angles $\beta_i$ ($1 \leq i \leq 4$). Eq. 8a should be replaced with Eq. 8b when $\beta_2 \leq \beta \leq \beta_3$ and $\beta_4 \leq \beta \leq \beta_1$.

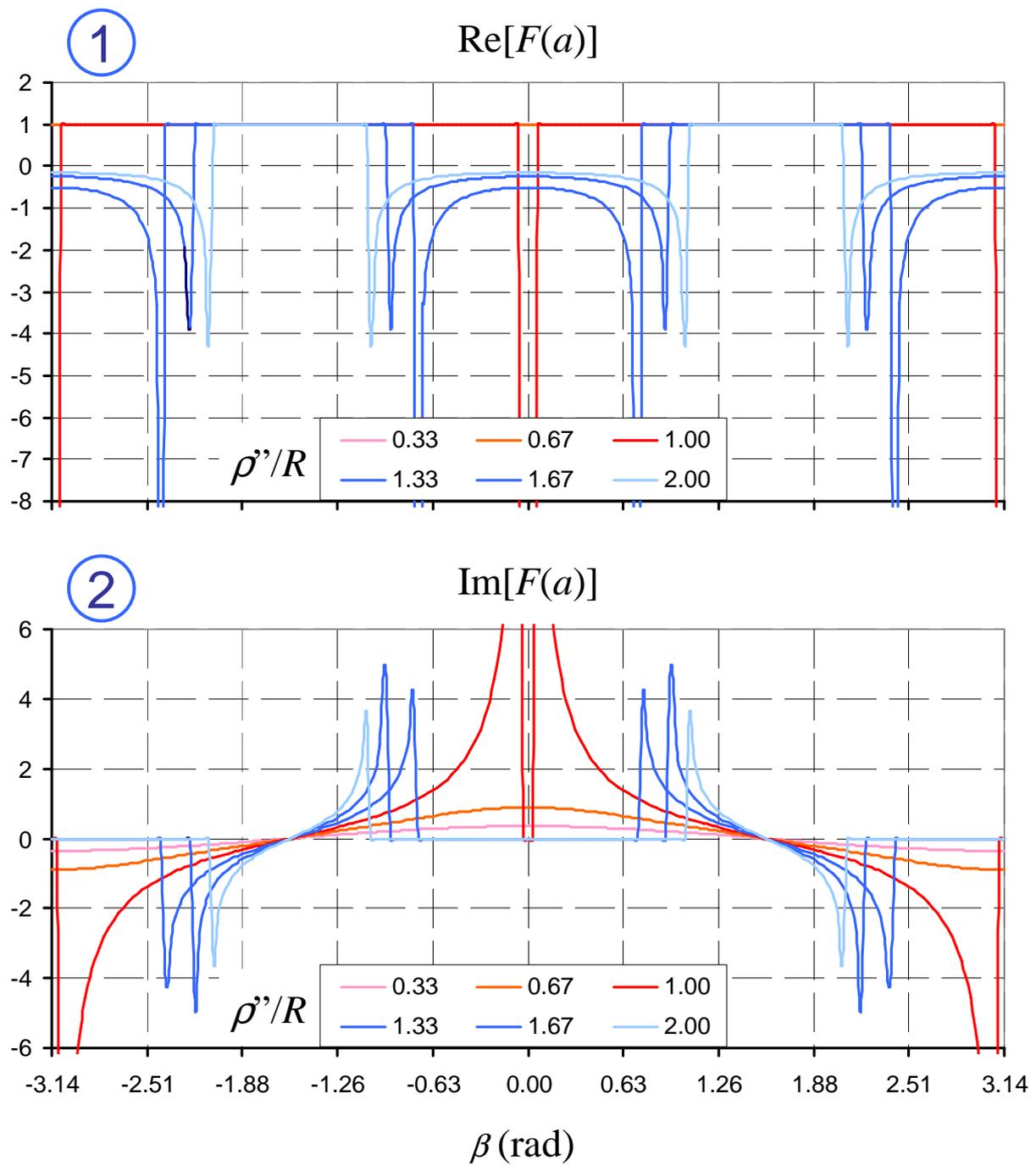

**Figure 5:** Plots of the real (1) and imaginary parts (2) of the weighting function $F(a)$ as function of the angle $\beta$, for different values of the ratio $\rho''/R$.

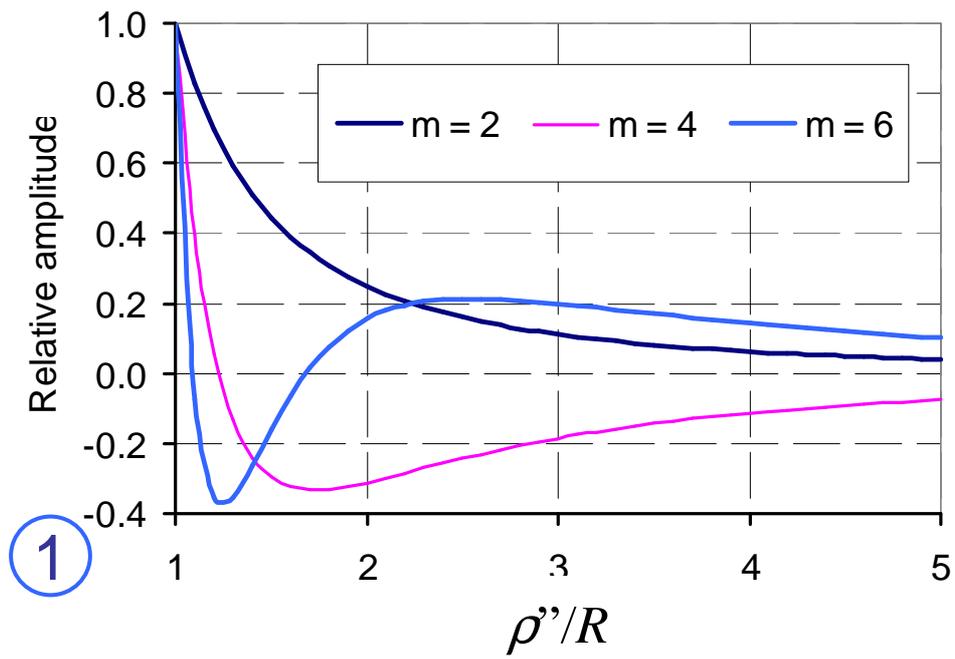

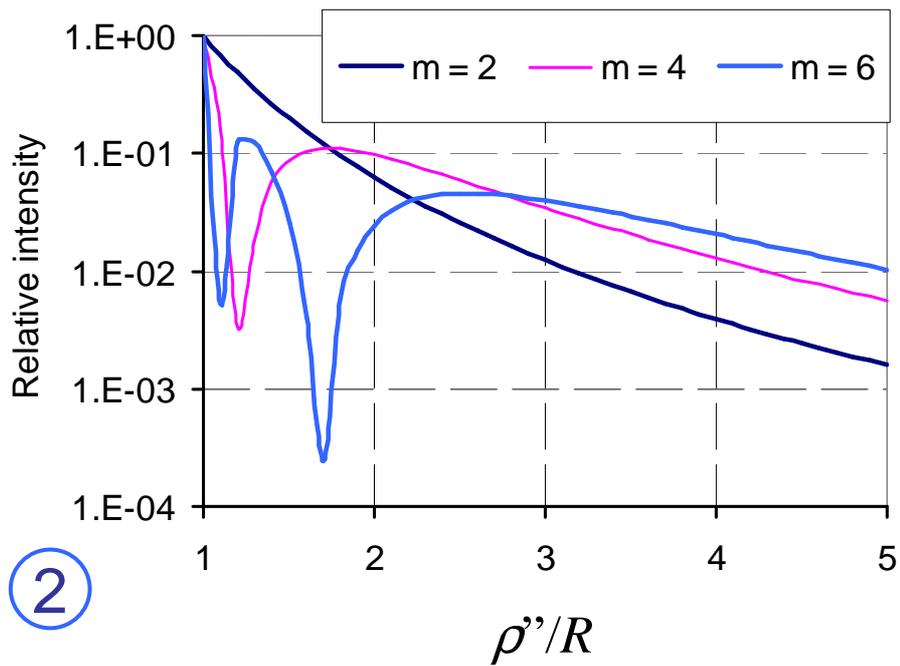

**Figure 6:** Amplitudes (1) and intensities (2) radiated outside of the telescope pupil by VPMs of topological charges $m$ = 2, 4 and 6, as function of the ratio $\rho''/R$. Intensities are plotted in logarithmic scale.

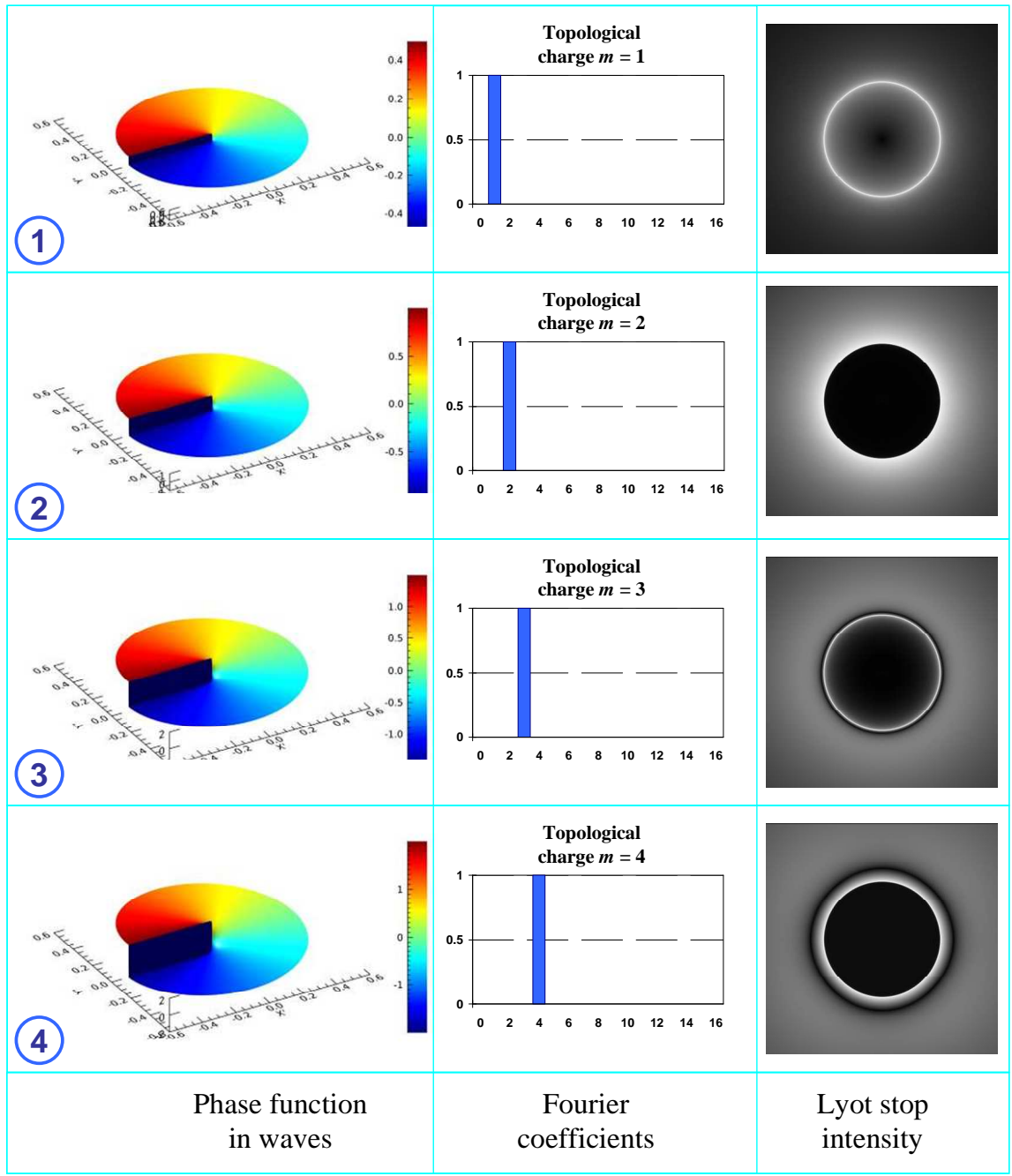

**Figure 7: Illustrating vortex phase functions of topological charges *m* = 1, 2, 3 and 4 (Figs. 7-1 to 7-4 respectively). For each case are shown three-dimensional plots of the phase function $\phi'(\theta')$ (left column), the modules of the Fourier coefficients $a_n$ defined in Eqs. 10 (central column), and grey-scale maps of the intensity distribution $I''(\rho'', \theta'')$ generated in the Lyot stop plane (right column). The latter are shown in logarithmic scale for enhancing the central extinction area.**

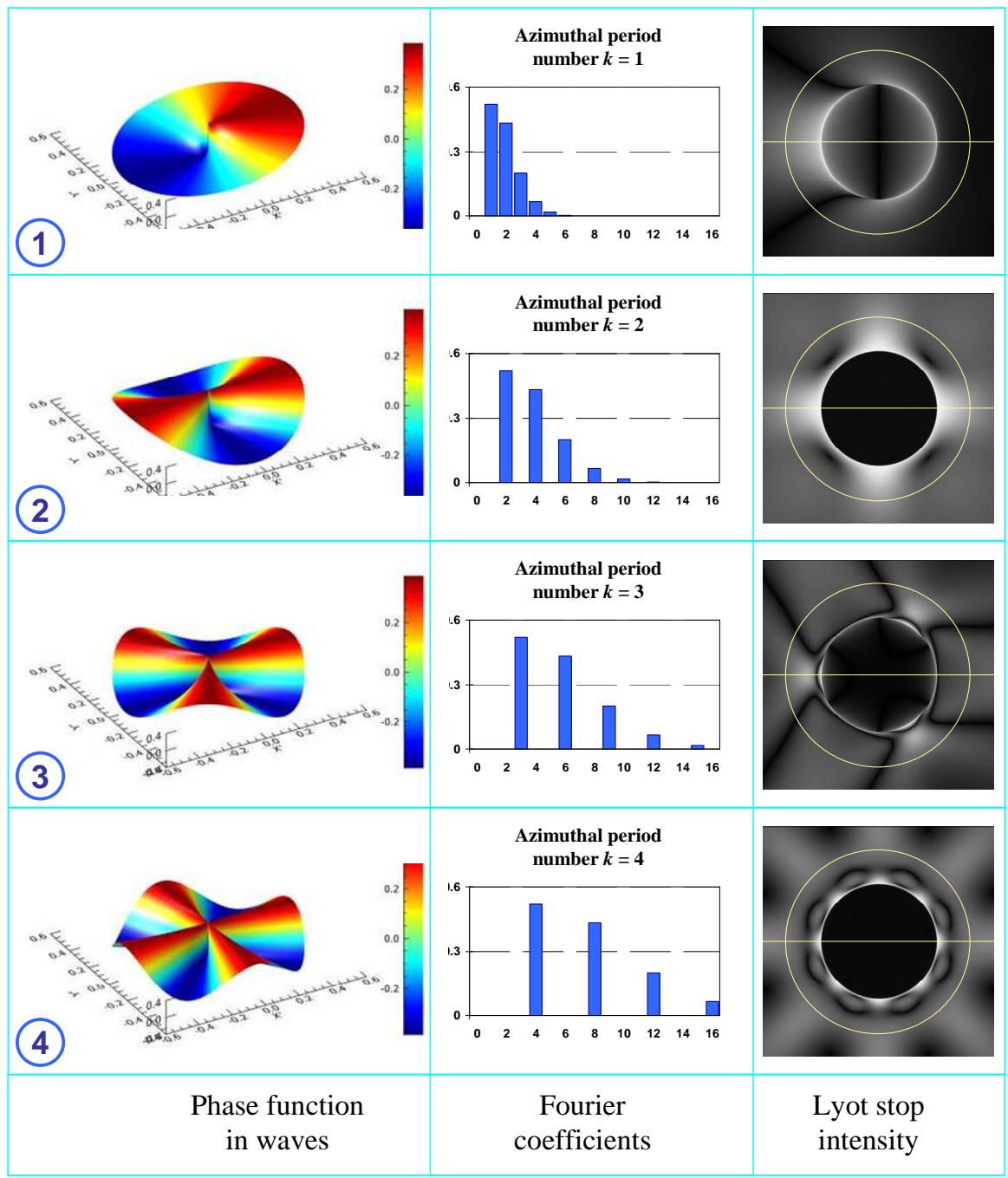

**Figure 8: Illustrating azimuthal cosine phase functions. Figs. 8-1 to 8-4 correspond to the cases of azimuthal period numbers $k$ = 1, 2, 3 and 4 respectively. Plots are similar to those of Figure 7, with white lines indicating the radial and azimuthal profiles of Figure 9.**

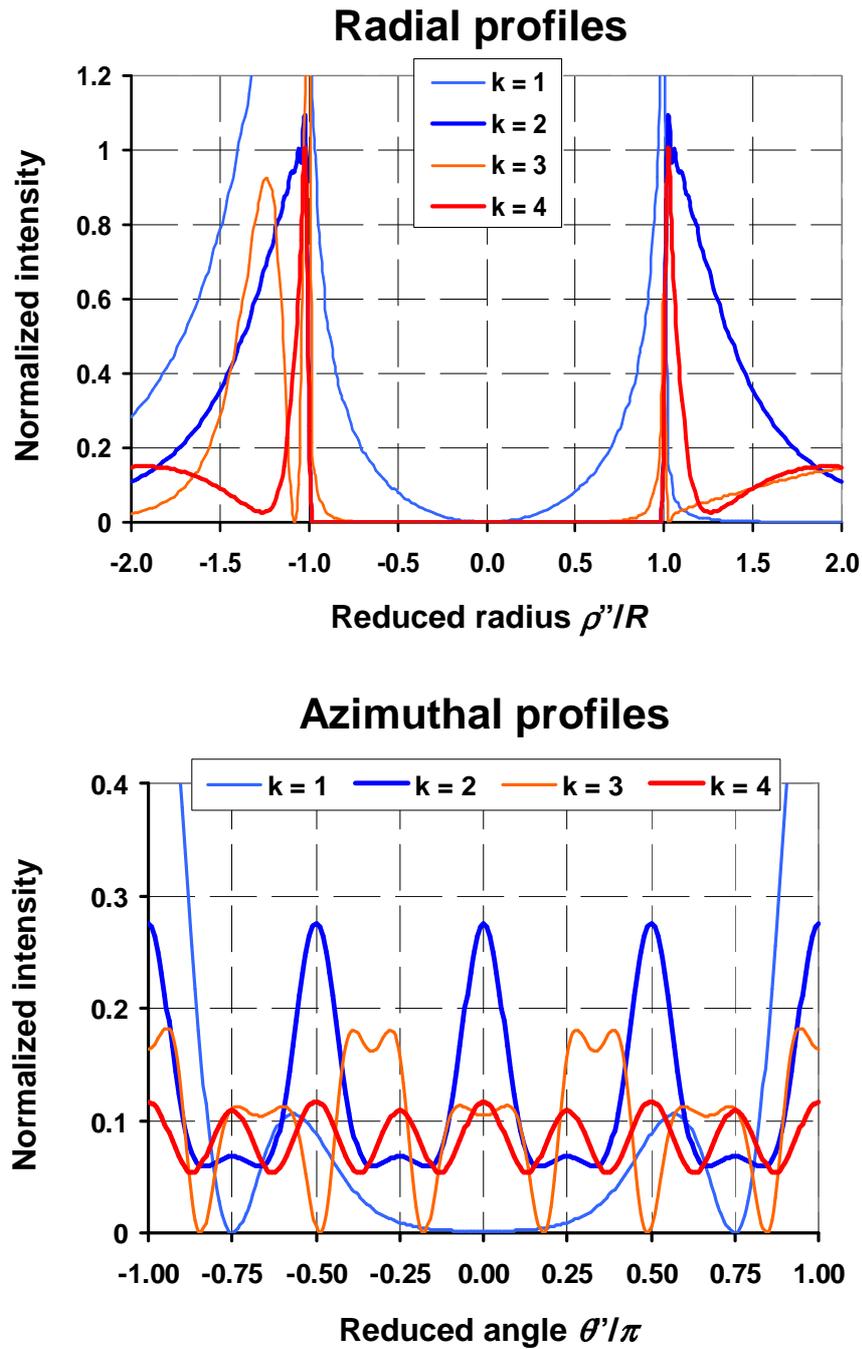

**Figure 9: Radial and azimuthal profiles of the intensity distributions generated by the ACM functions of Figure 8 into the Lyot stop plane. Top: Radial profiles at angle $\theta' = 0$ degree. Bottom: Azimuthal profiles at radius $\rho'' = 1.6\,R$.**

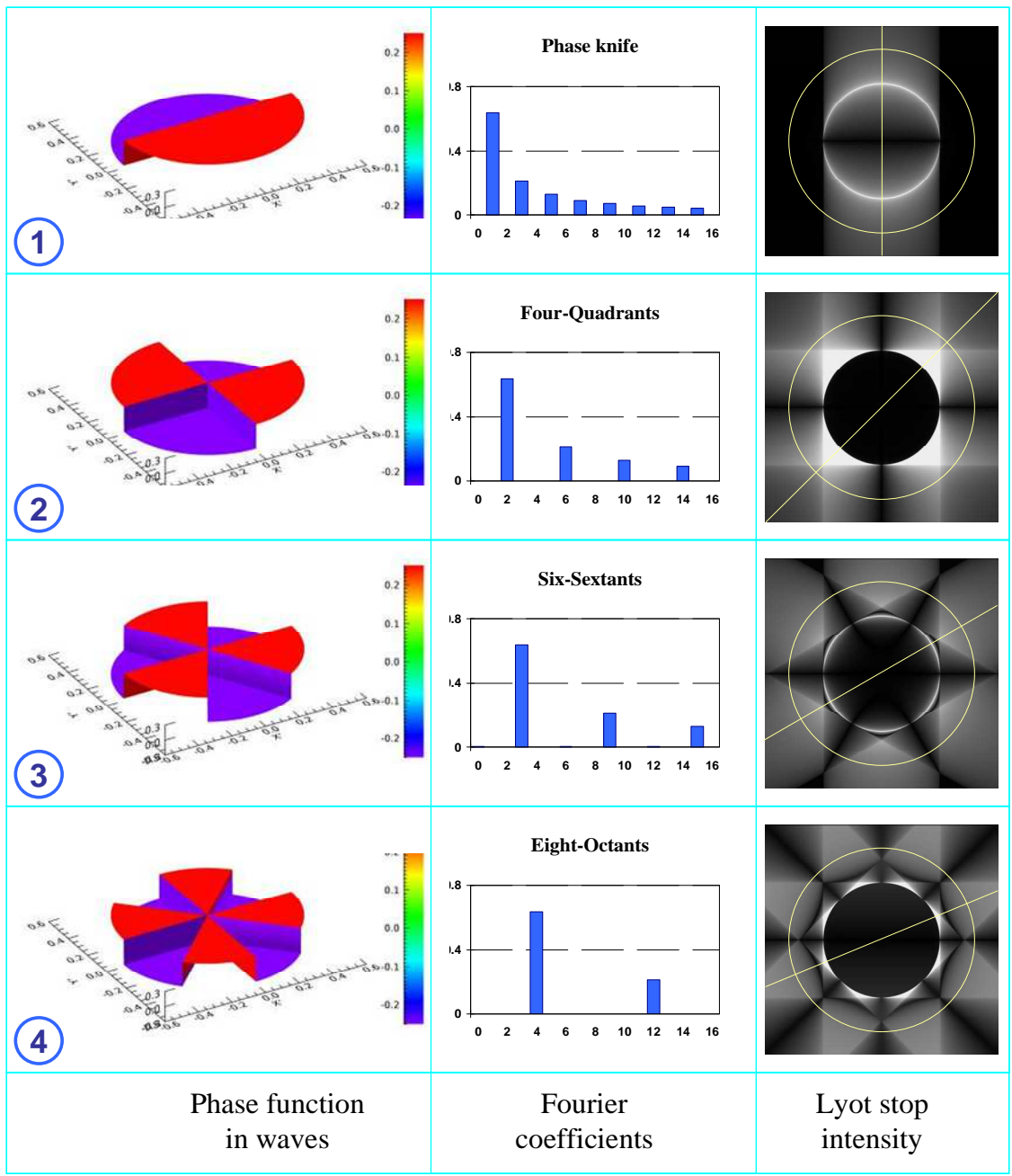

**Figure 10: Illustrating the case of regularly segmented phase functions. Figs. 9-1 to 9-4 respectively show the phase knife, four-quadrants, six-sextants and eight-octants phase masks. Plots are similar to those of Figure 7. White lines indicate the radial and azimuthal profiles of Figure 11.**

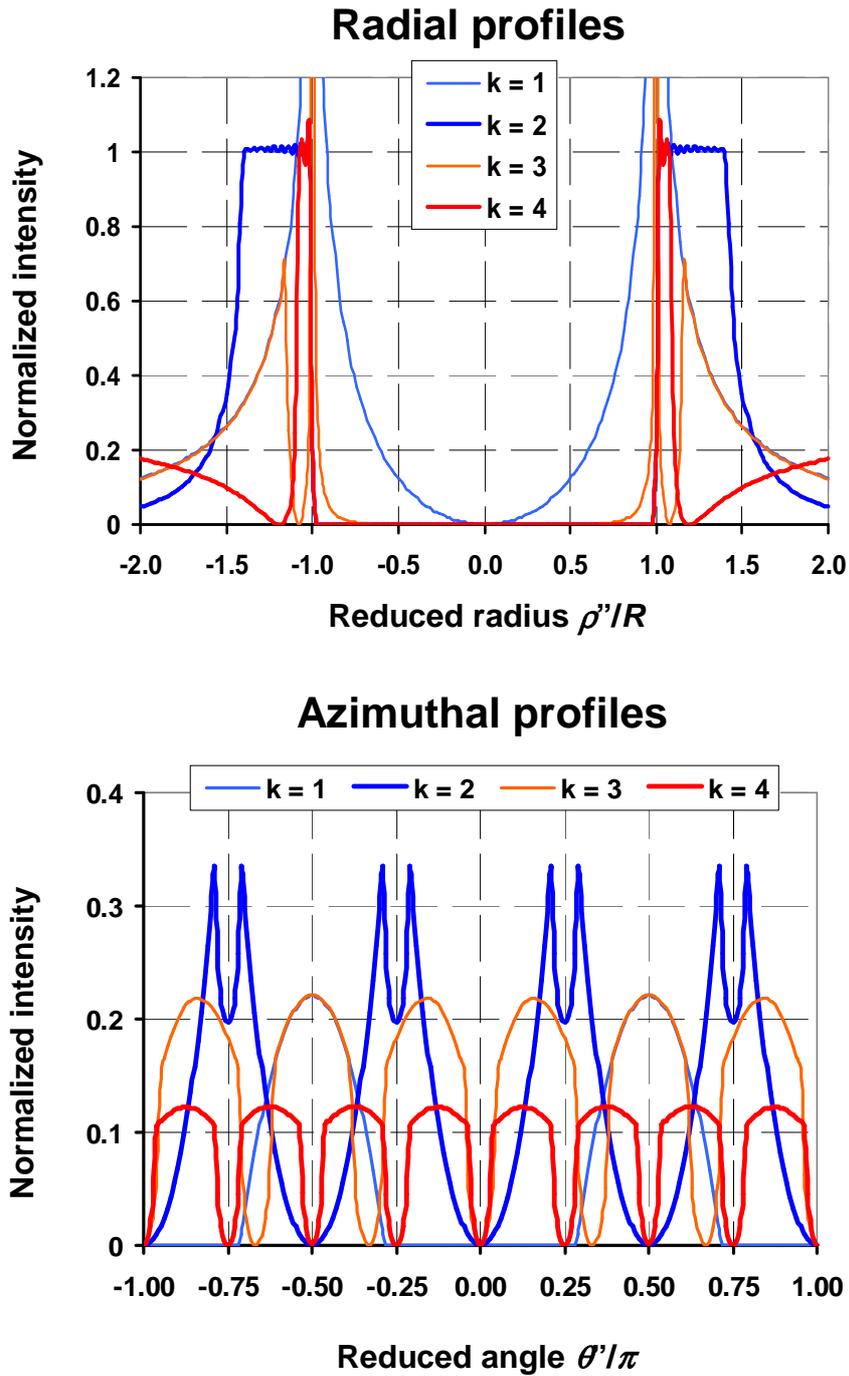

Figure 11: Radial and azimuthal profiles of intensity distributions generated by the segmented phase functions of Figure 10. Top: Radial profiles at angles $\theta'$ = 90; 45, 30 and 22.5 degrees for $k$ = 1, 2, 3 and 4 respectively. Bottom: Azimuthal profiles at $\rho'$ = 1.6 $R$.

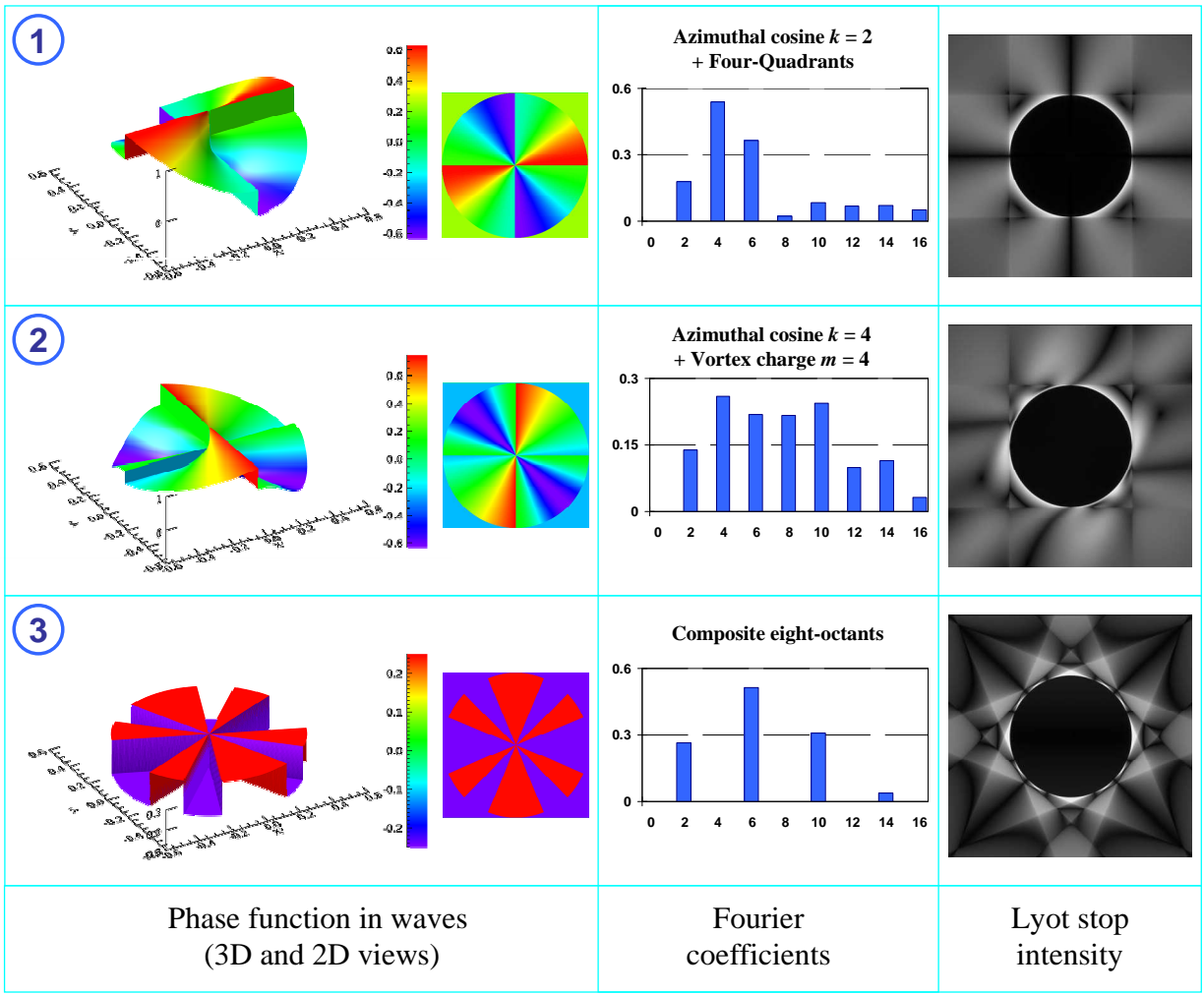

**Figure 12: Illustrating the case of regularly segmented phase functions. Plots are similar to those of Figure 7-9.**

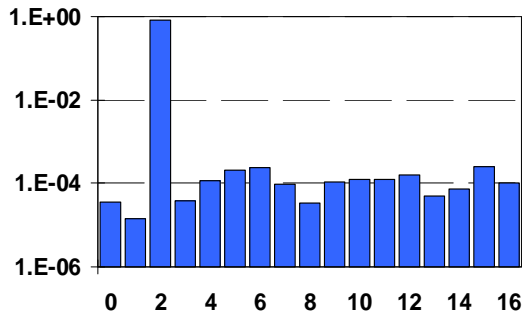 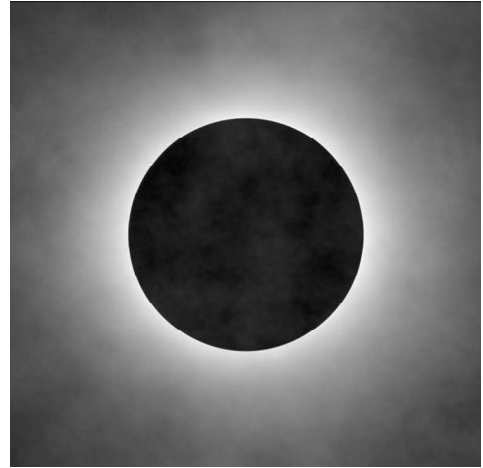

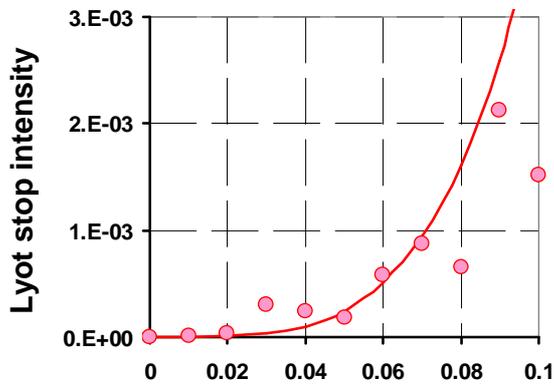 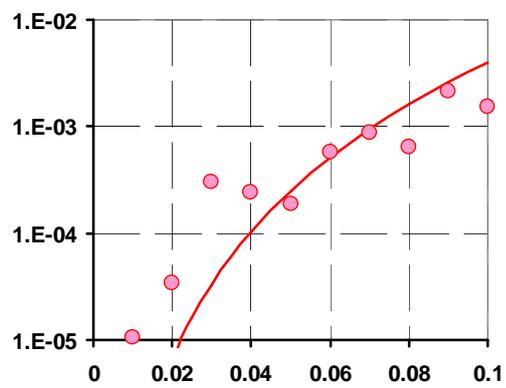

**Figure 13: Introduction of phase errors into a vortex mask of charge *m* = 2. (1) Fourier coefficients. (2) Intensity distribution in Lyot stop plane . Both plots are shown in logarithmic scale. (3) and (4): Curves of the mean irradiance inside the Lyot stop as function of phase errors standard deviation, in linear and logarithmic scale respectively. Numerical simulation results are indicated by red dots and the analytical formula 27 by red solid lines.**

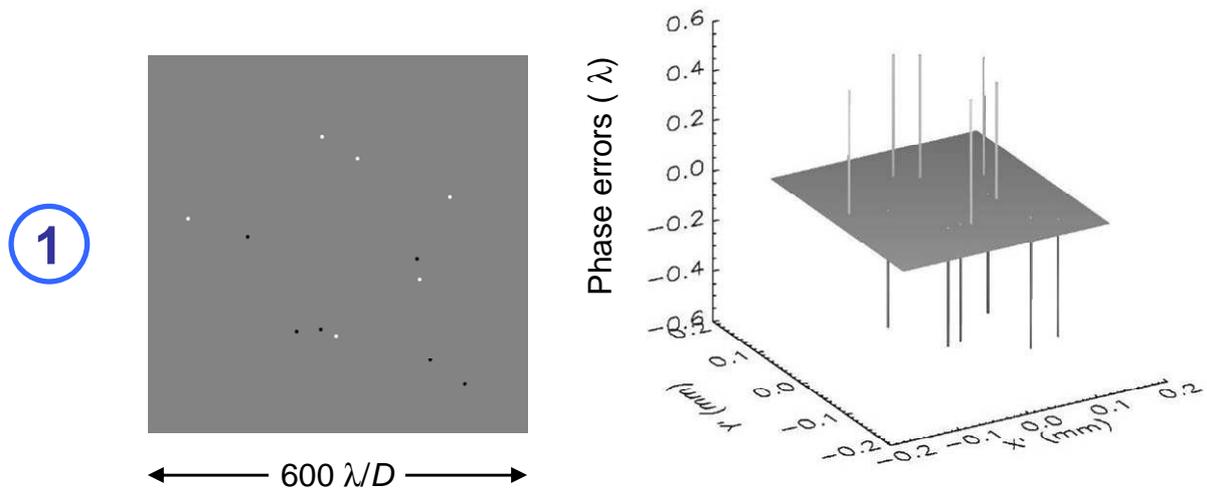

**Phase mask errors**

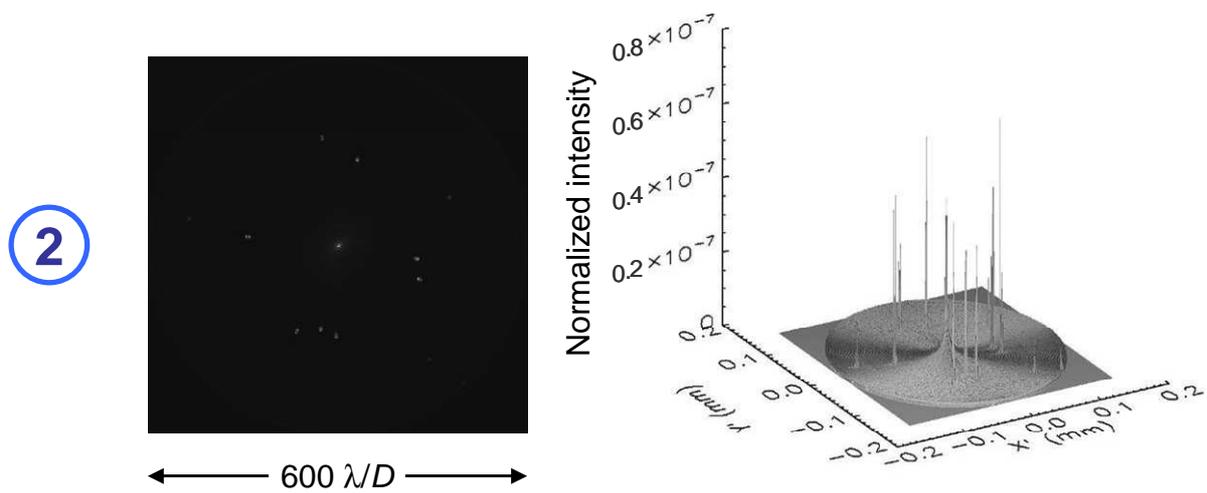

**Intensity in coronagraph image plane**

Figure 14: Introduction of local phase errors over a vortex mask of charge $m = 2$. (1) Grey-scale map and three-dimensional plot of input phase errors. (2) Same illustrations of intensity distribution generated in the coronagraph image plane. Local errors lay inside a disk of diameter 0.3 mm in the phase mask plane, which is equivalent to $600\lambda/D$.